\documentstyle[11pt,a4,cite]{article}

\def\ga{\alpha}

\def\ge{\epsilon}
\def\gg{\gamma}
\def\gd{\delta}
\def\gf{\phi}
\def\gm{\mu}
\def\gn{\nu}
\def\gp{\pi}
\def\gP{\Pi}
\def\gff{\varphi}

\def\gch{\chi}

\def\gL{\Lambda}
\def\gt{\theta}

\def\gz{\zeta}
\def\delp{\partial_+}
\def\delm{\partial_-}
\def\part{\partial}
\def\hlf{\frac{1}{2}}
\def\A0{A^{+}_0}
\def\psip{\psi_+}
\def\psin{\psi_-}
\def\psib{\overline{\psi}}

\def\xmin{x^{-}}
\def\ymin{y^{-}}
\def\xpl{x^+}
\newcommand{\nc}{\newcommand}
\nc{\intgl}{\int\limits_{-L}^{+L}\!{dx^-\over 2}}
\nc{\intgly}{\int\limits_{-L}^{+L}\!{dy^-\over 2}}      
\nc{\zmint}{\int\limits_{-L}^{+L}\!{{dx^-}\over{\!2L}}}
\def\beq{\begin{equation}}
\def\eeq{\end{equation}}
\def\bea{\begin{eqnarray}}
\def\eea{\end{eqnarray}}
\begin{document}
\title{Theta - Vacuum of the Bosonized Massive Light--Front Schwinger Model}
\author{{\sl L$\!$'ubom\'{\i}r Martinovi\v c} \\
Institute of Physics, Slovak Academy of Sciences \\
D\'ubravsk\'a cesta 9, 842 28 Bratislava, Slovakia \thanks{permanent address}\\
and\\
International Institute of Theoretical and Applied Physics,\\
\medskip
Iowa State University, Ames, Iowa 50011, USA \\ 
{\sl James P. Vary}  \\
Department of Physics and Astronomy \\
and \\ 
International Institute of Theoretical and Applied Physics, \\
Iowa State University, Ames, Iowa 50011, USA}
\date{October 5, 1998}
\maketitle
\begin{abstract}

The massive Schwinger model in bosonic representation is quantized on the 
light front using the Dirac--Bergmann method. The non-perturbative theta 
vacuum in terms of coherent states of the gauge-field zero mode is derived  
and found to coincide with the massless case. On the other hand, the mass 
term becomes highly non-linear due to the constrained zero mode of the scalar 
field. A non-trivial mixing between the normal-mode and zero-mode sectors of 
the model is crucial for the correct calculation of the theta dependence of 
the leading order mass correction to the chiral condensate. 
\end{abstract}
%\pacs{ }

\section{Introduction}

Massless quantum electrodynamics in two dimensions \cite{Schw}, the Schwinger  
model, is a solvable model with a surprising richness of non-perturbative 
phenomena. It has become a useful testing ground for new methods in quantum 
field theory including the method of discretized light-cone quantization (DLCQ) 
\cite{EPB}. Full mass spectra with the corresponding wave functions have been 
obtained for broad ranges of the coupling constant (and the fermion mass). It 
turns out however that it is much more difficult to understand such  
non-perturbative aspects known from usual formulation as chiral anomaly, 
$\theta$ vacuum and chiral symmetry breaking within DLCQ \cite{Rgf}. It is even   
not quite clear which degrees of freedom are responsible for these phenomena.    
 
An attempt to derive a full operator solution of the light-front Schwinger   
model has been undertaken by McCartor \cite{McC} (see also \cite{McCPR}). An 
important ingredient in this approach was an initialization of quantum fields on  
the both characteristic surfaces $x^+=0, x^-=0$. Consequently, some advantages  
of the light-front (LF) formulation (first of all the ``triviality'' of the LF  
vacuum) have been lost. A genuine light-front solution equivalent to that of  
the covariant formulation \cite{LSw,MPS} has not been given so far. 

On the other hand, some progress in understanding non-perturbative aspects 
of LF field theories has been achieved by studying the constrained and 
dynamical zero modes of bosonic fields. They appear due to periodic boundary 
conditions in the finite-volume formulation of the theory \cite{MY,Franke}. 
In particular, the dynamical zero mode of the $A^+$ gauge-field component 
\cite{Rgb} has been shown to lead to a non-trivial vacuum structure in the 
bosonized (massless) Schwinger model within the Schr\"odinger coordinate 
representation \cite{AD} as well as in the coherent state approach in the  
Fock representation \cite{lm1}. 

In addition, improved light-front gaussian effective potential methods 
\cite{PNPir} and near light-front quantization approaches \cite{VFPir} have 
provided access to the condensate and chiral corrections.

The purpose of the present paper is to give a generalization of the 
coherent-state approach \cite{lm1} for the case of non-vanishing mass of 
the fermion field. The Lagrangian of the bosonized theory is employed for 
simplicity. This helps to avoid so far unsolved problems with massless LF 
fermions in two dimensions \cite{Rgf,McC} while still allows to study some 
non-trivial aspects of the LF vacuum structure. We choose the Dirac--Bergmann 
quantization method to correctly handle the constraints present in both the 
normal and zero Fourier mode sectors. The procedure generates a secondary 
constraint which relates the zero mode of the scalar field and the gauge field 
conjugate momentum. As a consequence, the bosonized fermion mass term becomes 
a complicated non-linear function of zero-mode variables. This can be 
understood as the price of constructing the exact ground state of the theory. 

Based on the quantum realization of the residual symmetry of the gauge zero 
mode under constant shifts, the true physical (gauge invariant) vacuum is 
indeed derived in terms of coherent states of the dynamical gauge field.  
The validity of the theta-vacuum construction is demonstrated by a calculation 
of the lowest order fermion-mass correction to the chiral condensate. The 
correct treatment of the commutators between the zero and normal modes of the 
scalar field plays a crucial role here.

\section{Bosonized massive LF Schwinger model in a finite box} 
 
Aside from a variational treatment \cite{HV}, the only known method to study 
properties of the vacuum state of two-dimensional LF field theories is to 
restrict the system to a finite interval $-L \leq x^{-} \leq L$. This can be 
viewed as a convenient infrared regularization which facilitates disentangling 
the vacuum aspects ($k^+=0$ modes of quantum fields) from the remainder of the 
dynamics. The decomposition of the fields into the zero-mode (ZM, subscript 
$0$) and normal-mode (NM, subscript n) parts can easily be performed by 
imposing periodic boundary conditions on the corresponding fields 
\footnote{The necessity to prescribe (quasi)periodic boundary conditions for  
the consistency of the LF quantization even in continuum formulation has been 
emphasized by Steinhardt \cite{Steinh}.}. For example, the gauge and scalar 
fields, whose dynamics will be studied below, are decomposed as (our 
convention is $x^\pm = x^0 \pm x^1$) 
\beq   
A^{\gm}(x^+,x^-)=
A^{\gm}_{0}(x^+)+A^{\gm}_{n}(x^+,x^-),\;\; \gf(x^+,x^-)=\gf_{0}(x^+)+\gf_{n}
(x^+,x^-).
\eeq
Before proceeding with the Dirac-Bergmann (DB) quantization procedure, 
let us recall the relation between the fermionic and bosonic formulations of 
the massive Schwinger model. The Lagrangian density of the two-dimensional  
spinor field $\psi$ of mass m interacting with the gauge field $A^\gm$ is 
\beq
{\cal L}  =  - \textstyle {1 \over 4} F^{\mu \nu} F_{\mu \nu} 
 + \bar{\psi} i \gamma^\mu D_\mu \psi - m\bar{\psi}\psi.
\label{covlagr}
\eeq 
In the light-front formulation with the finite-volume light cone gauge 
$A^{+}_{n}=0, A^{-}_{0}=0$ we have 
%\cite{Col,Mand}
\bea
{\cal L}_f & = & 2i\psip^{\dagger}\delp\psip + 2i\psin^{\dagger}\delm\psin + 
{\hlf}(\delp\A0)^2
 + {\hlf}(\delm A^{-}_{n})^2 \nonumber \\ 
& - & m(\psip^{\dagger}\psin + \psin^{\dagger}\psip) 
- {e \over 2}j^{+}_{n}A^{-}_{n} - {e \over 2}j^{-}_{0}\A0 .
\label{flagr}
\eea 
Here $\psip$ and $\psin$ are respectively the dynamical and dependent 
fermi-field components 
\beq
\psip=\gL_+\psi,\;\psin=\gL_-\psi,\;\gL_{\pm}=\hlf\gg^0\gg^{\pm},\;\gg^{\pm}
=\gg^0\pm \gg^1\;.
\eeq
The bosonized form of the theory is obtained by the correspondences 
\cite{KSus,Col}
\beq
i\psib\gg^\gm\part_\gm \psi = \hlf \part^\gm \gf \part_\gm \gf,
\label{br1}
\eeq
\beq
j^{\gm}={1 \over \sqrt{\gp}}\ge^{\gm \gn}\part_{\gn} \gf,
\label{br2}
\eeq
\beq
\psib \psi=K:\cos c\gf:,
\label{br3}
\eeq
where $\gf(x)$ is the equivalent real boson field. The other symbols in 
(\ref{br1}) -- (\ref{br3}) are
\beq
K={\gm \over {2\gp}}e^{\gg_{E}}\;,\;\;\;\gm={e \over{\sqrt{\gp}}},\;\; 
\ge^{\gm \gn} = -\ge^{\gn \gm},\;\;c=2\sqrt{\gp} ;
\eeq
$\gm$ is the Schwinger boson mass and $\gg_E$ is Euler's constant.
%j^{\gm}_{5}={1 \over \sqrt{\gp}}\part^{\gm} \gf\;,\;\;\;
The bosonized Lagrangian density in the chosen gauge has the form 
\bea
{\cal L}_{b} & = & 2\delp \gf_n \delm \gf_n +{\hlf}(\delp \A0)^2 + 
\hlf (\delm A^-_n)^2 - \gm A^-_n \delm\gf_n  \\
 & + & \gm\A0\delp \gf_0 - mK:\cos \left(c\gf_n + c\gf_0 \right):.
\label{blagr}
\eea 
%\beq
%{\cal L} = {\hlf}(\part_{\gm} \gf)(\part^{\gm} \gf)-{1 \over 4}
%F_{\gm \gn}F^{\gm \gn}\;-\;ej_{\gm} A^{\gm} - mK:\cos 2\sqrt{\gp} \gf:, 
%\label{blagr} 
%\eeq
Note that one had to fix the gauge already in the fermionic Lagrangian (\ref
{flagr}) \cite{KSus}, because the equivalent form of the fermion kinetic energy  
is (trivially) gauge invariant by itself and does not compensate the gauge 
variation of the interacting term. In other words, it is difficult to speak 
about gauge symmetry in the bosonized form of the Lagrangian obtained   
by the covariant bosonization rules (\ref{br1}) -- (\ref{br3}). The sole 
remnant of the original gauge freedom is the symmetry under constant shifts of 
$\A0$, which is `visible' only in the finite-box formulation. This symmetry  
under ``large" gauge transformations will emerge as the source of the  
non-trivial vacuum structure of the model. 

Proceeding with the DB procedure, we compute momenta conjugate to the fields 
in the Lagrangian (\ref{blagr}):  
\beq
\gP_{A^-_n} = 0,\;\;\;
\gP_{\gf_{n}}  =  2\delm \gf_{n},\;\;\; 
\gP_{\A0} = \delp \A0,\;\;\;
\gP_{\gf_{0}}  = \gm \A0. 
\eeq
The momenta, which contain no $\xpl$-derivative, give rise to the NM and ZM 
primary constraints 
\beq
\gff_1 = \gP_{A^-_n},\;\;\;\gff_2=\gP_{\gf_{n}}-2\delm\gf_{n},\;\;\;
\gff_3=\gP_{\gf_{0}}-\gm\A0.
\label{primcon}
\eeq
The primary Hamiltonian is         
\beq
P^{-}_{p}=P^{-}_{c} + Lu_3 \gff_3 + \intgl \left[ u_1(x) \gff_1(x) + 
u_2(x) \gff_2(x)\right]
\label{primham}
\eeq
with the canonical LF Hamiltonian $P^{-}_{c}$
\beq
P^{-}_{c} = \intgl \left[\gP^{2}_{\A0} - (\delm A^-_n)^2 +  
 2 \gm A^-_n \delm \gf_n + 2mK:\cos c(\gf_{0}+\gf_{n}): \right]
\label{lfham}
\eeq
derived from the Lagrangian (\ref{flagr}) in the standard way. One has to check 
$\xpl$-independence of the primary constraints next. Consistency of $\gff_1$ 
and $\gff_3$ generates secondary constraints 
\bea 
\gch_1 & = & \delm A^-_n + \gm \gf_n, \\ 
\gch_2 & = & \gP_{\A0} - {{mKc}\over \gm}\left(:\sin c\gf:\right)_0. 
\label{seccon}
\eea
(The subscript $0$ indicates the ZM projection of the expression in 
the parenthesis.) Consistency of $\gff_2$ as well as of $\gch_1$ and $\gch_2$ 
yields weak equations for the Lagrange multipliers $u_2, u_1, u_3$, which 
means that this part of the procedure terminates. Since the freedom under 
small gauge transformations has been removed at the Lagrangian level, all five 
constraints (in the order $\gff_1,\gff_2,\gch_1,\gff_3,\gch_2)$ are second 
class and can be used to calculate the matrix of their Poisson brackets. Its  
inverse, needed for the computation of the Dirac brackets, is 
\beq
C^{-1}(\xmin - \ymin) = \left({\matrix{{\!\gm^2\over 4}{\cal G}_3 & \!
{\cal G}_1 & \! {\gm \over 4}{\cal G}_2 & \! 
C^{-1}_{14} & \! 0 \cr
{\! \cal G}_1 &\! 0 & \!0 & \!0 & 0 \cr \! -{\gm \over 4}
{\cal G}_2 & \!
0 & \!-{1 \over 4}{\cal G}_1 & \!C^{-1}_{34} & \! 0 \cr 
\!C^{-1}_{41} & \!0 & \! C^{-1}_{43} & \! 0 & {L \over 
\gm}{1 \over{1 - \ga_0}} \cr \! 0 & \!0 & \!0 & \!-{L\over \gm}
{1 \over {1 - \ga_0}} & \! 0 }} \right) ,
\label{inversec}
\eeq
where the argument of the matrix elements has been suppressed.
The NM Green's functions ${\cal G}_k(\xmin - \ymin)$ are defined by the 
equations ($k=1,2,3$, no summation)
\beq
\delm^k {\cal G}_k(\xmin - \ymin) = \gd_n(\xmin - \ymin), 
\eeq
where $\gd_n(\xmin)$ and ${\cal G}_1(\xmin)$, etc. are the NM parts of the   
periodic delta function and one half of the periodic sign function 
$\ge_n(\xmin)$, respectively \cite{AC}. 
The symbolic matrix elements in $C^{-1}$ have the following explicit form: 
\bea
C^{-1}_{14}(\xmin) = {mKc^2 \over{4 \gm (1-\ga_0)}}
\intgly {\cal G}_2(\xmin - \ymin):\cos c\gf(\ymin): \nonumber \\
C^{-1}_{34}(\xmin) = {mKc^2 \over {4 \gm^2 (1 - \ga_0)}}
\intgly {\cal G}_1(\xmin - \ymin):\cos c\gf(\ymin): ,
\eea
and similarly for $C^{-1}_{41}(\ymin), C^{-1}_{43}(\ymin)$. The quantity 
$\ga_0$ is the ZM projection of 
\beq
\ga(\xmin) = {mKc^2 \over {\gm^2}}:\cos c\gf(\xmin): .
\eeq
The above $C^{-1}_{ij}$ belong to the mixed ZM/NM sector of the matrix 
$C^{-1}$, while the lower right 2 by 2 submatrix corresponds to the ZM sector. 

We do not quote all calculated Dirac brackets here. In the NM sector the only 
relevant commutator for our purpose in the quantum theory is
\beq
%\left[A^-_n(\xmin),\gf_n(\ymin)\right] = i{\gm\over 4}{\cal G}_2(\xmin-\ymin),
\left[\gf_n(\xmin),\gf_n(\ymin)\right] = -{i \over 4}\hlf\ge_n(\xmin-\ymin).
\eeq
The rest of the commutators can be obtained by differentiation using the 
corresponding constraint (\ref{primcon}) strongly. The Dirac brackets, signified by an 
asterisk, in the ZM sector have a more complicated structure. The non-
vanishing cases are 
\bea
\{\A0,\gP_{\A0}\}^* & = & {1 \over L}(1 - {1 \over{1 - \ga_0}}),\;\;\;
\{\A0,\gf_0\}^*  =  -{1 \over{L\gm}}{1 \over {1 - \ga_0}},\nonumber \\
\{\gP_{\A0},\gP_{\gf_0}\}^* & = & {\gm \over L}{\ga_0 \over{1 - \ga_0}},\qquad
\;\;\;\;\{\gf_0,\gP_{\gf_0}\}^*  =  {1 \over L}{1 \over{1 - \ga_0}} .
\eea
These can be simplified by defining new variables
\beq
\gP^-_0 =  \gP_{\A0} - \gm\gf_0,\;\;\;\;\gP_0 = \gP_{\gf_0} - \gm\A0.
\label{vchange}
\eeq
The only non-zero commutators in quantum theory then read
\beq
\left[\A0,\gP^-_0 \right] = {i\over L},\;\;\;\;\left[\A0,\gf_0 \right] = 
-{i \over 
{\gm L}}{1 \over {1 - \ga_0}} .
\label{zmcr}
\eeq
Note, that in the massless limit $m = 0$, $\gP^-_0$ becomes $-\gm\gf_0$ and 
both commutators coincide in agreement with direct $m=0$ calculations \cite{AD,
lm1}. Finally, there is one independent non-zero mixed commutator 
\beq
\left[\gf_n(\xmin),\gf_0 \right] = -i{mKc^2 \over {4\gm L(1 - \ga_0)}} 
\intgly {\cal G}_1(\xmin - \ymin):\cos c\gf(\ymin): .
\label{mixedcr}
\eeq
The quantum LF Hamiltonian, which depends only on unconstrained field 
variables, is obtained by inserting the constraints $\gch_1$ and $\gch_2$ 
into the primary Hamiltonian (\ref{lfham}) as strong operator relations: 
\beq
P^- = \left[{mKc \over \gm} \left(:\sin c\gf:\right)_0\right]^2 + \intgl 
\left[\gm^2 \gf^2_n(\xmin) + 2mK:\cos c\gf(\xmin): \right].
\label{qham}
\eeq  
This completes the LF Hamiltonian quantization of the massive Schwinger model.

\section{Residual gauge symmetry and the $\gt$-vacuum}

Even after the complete gauge fixing at the classical level, the Lagrangian 
(\ref{flagr}) has a residual large gauge symmetry characterized by the   
gauge function, linear in $\xmin$ 
\beq
\gL_\gn =  {\gp \over {eL}}\gn\xmin,\;\;\gn \in Z,  
\eeq
which tends to a non-zero constant at $\xmin = \pm L$. The linearity in 
$\xmin$ and the combination of constants in the coefficient are the 
consequence of a requirement to maintain boundary conditions for the 
gauge and fermi fields, respectively. 

As discussed above, the bosonization rules (\ref{br1}) -- (\ref{br3}) can only be  
applied in the gauge-fixed situation. The persistant part of the original symmetry  
in the bosonized LF Hamiltonian (\ref{qham}) consists of constant shifts of $\A0$ 
\cite{AD}:  
\beq
\A0 \rightarrow \A0 - {2\gp \over {eL}}\gn. 
\label{ashift}
\eeq 
Although $P^-$ does not explicitly dependend on $\A0$, an important 
piece of information carried by this gauge degree of freedom is encoded in the 
first commutation relation in (\ref{zmcr}). The latter can be used to 
eliminate at the state vector level the arbitrariness related to the constant 
shifts (\ref{ashift}). To do this, it is helpful to define the rescaled ZM  
variables \cite{AD} 
\beq
\A0 = {2\gp \over{eL}}\hat{\gz},\;\;\gP^-_0 = {e \over {2\gp}}\hat{\gp}_0,
\eeq
in terms of which the basic commutation relation takes a simple quantum-
mechanical form, independent of the box length L:
\beq
\left[\hat{\gz},\hat{\gp}_0\right] = i. 
\eeq
It is simple to see that quantum-mechanically the shift transformation 
\beq
\hat{\gz} \rightarrow \hat{\gz} - \gn
\eeq
is realized by the unitary operator $\hat{T}_\gn = (\hat{T}_1)^\gn$
\beq
\hat{\gz} \rightarrow \hat{T}_\gn \hat{\gz} \hat{T}^{\dagger}_\gn,\;\;
\hat{T}_\gn = 
\exp(-i\gn \hat{\gp}_0).
\eeq
The same operator transforms also the vacuum state. In the coordinate 
representation, $\hat{\gp}_0$ and the vacuum are expressed as 
($a_0$ is the ZM annihilation operator)
\beq
\hat{\gp}_0 = -i {d \over {d\gz}},\;a_0\psi_0(\gz) \equiv (\gz + {d \over d\gz})
\psi_0(\gz) = 0,\;\;\psi_0(\gz) = {\gp}^{-{1 \over 4}} \exp\left(-\hlf 
\gz^2\right). 
\label{vacdef}
\eeq
Then the displacement operator $\hat{T}_1$ acts simply on the trivial vacuum 
$\psi_0(\gz)$:
\beq
\psi_0(\gz) \rightarrow \hat{T}_\gn \psi_0(\gz) = \psi_\gn(\gz) = {\gp}^
{-{1 \over 4}} \exp \left(-\hlf(\gz - \gn)^2\right) .
\eeq
In this way, there are infinitely many degenerate vacuum states $\psi_\gn, 
\gn \in Z$, corresponding to the infinite set of shifted 
ZM variables $\A0$ (or $\gz$). The operator $\hat{T}_1$ acts as a raising 
operator and to have a vacuum state, invariant under $\hat{T}_1$, 
we need to superimpose all states to form the $\gt$-vacuum:
\beq
\vert \gt \rangle = \sum_{\gn=-\infty}^{\infty} e^{-i\gn\gt}\psi_
\gn(\gz)|0\rangle,
\eeq
($\vert 0 \rangle$ is the vacuum with respect to the $\gf_n(\xmin)$ field) 
with the desired propery - invariance up to a phase
\beq
\hat{T}_1 \vert \gt \rangle = e^{i\gt} \vert \gt \rangle .
\label{phaseinv}
\eeq
Thus, that part of the original gauge symmetry, which is not related to the 
redundant gauge degrees of freedom, gives rise -- when realized 
in accord with quantum mechanics -- to the multiple vacua. The requirement of  
gauge invariance of the true physical vacuum then implies existence of the $\gt
$-vacuum for the massive Schwinger model quantized on the light front. Due to 
bosonization, the structure of the coherent-state vacua $\psi_\gn(\gz)$ is 
fully described by only one gauge degree of freedom and is actually very 
simple. The same task will be harder within the original fermion 
representation, because one has to find a mechanism to enrich the vacuum by a 
fermion component. The latter is inevitable for a non-zero chiral condensate. 

In the present formulation, it is not too difficult to calculate even the 
$O(m)$ correction to the condensate. For this purpose, one notes that the ZM part of the scalar field in the 
bosonized form of the fermi field bilinear (\ref{br3}) is due to the 
definition (\ref{vchange}) expressed as
\beq
\gf_0 = {1 \over \gm}\gP_{\A0} - {1 \over \gm}\gP^-_0.
\label{subst}
\eeq
Using the rescaled momentum $\hat{\gp}_0$, one arrives at
\beq 
\psib \psi = K:\cos (c\gf_n + c\gf_0): = K :\cos \left(c\gf_n + 
{c \over \gm}\gP_{\A0} - \hat{\gp}_0 \right):.
\eeq
Recall however that $\gP_{\A0}$ is not an indpendent variable -- it has to obey 
the secondary constraint $\gch_2$ Eq.(\ref{seccon})
\beq
\gP_{\A0} = {mKc \over \gm} \left(:\sin (c\gf_n + c\gf_0):\right)_0,
\eeq
where the argument of $\sin$ on the r.h.s. itself contains $\gP_{\A0}$ through 
$\gf_0$ (\ref{subst}). It is thus evident that we are dealing with a highly 
non-linear problem which can only be solved approximately. A natural method 
is to iterate in the fermion mass $m$. Writing $\cos$ in the exponential form, 
one gets
\beq
:\cos c\gf: = \hlf\left[:\exp\left(ic\gf_n + i{c \over \gm}\gP^{(1)}_{\A0}
\right): \hat{T}_1 + h.c. \right],
\label{cos1}
\eeq
where 
\beq
\gP^{(1)}_{\A0} = {mKc \over \gm} \left(:\sin(c\gf_n - \hat{\gp}_0):\right)_0 
\eeq
is the lowest-order (in $m$) approximation of $\gP_{\A0}$. After expanding 
(\ref{cos1}) in $m$ and using the $\gt$-vacuum property (\ref{phaseinv}), 
one readily finds
\beq
\langle \gt \vert \psib\psi \vert \gt \rangle = K \langle \gt \vert :\cos c\gf :
\vert \gt \rangle = K \cos \gt - m {e^{2\gg_E} \over \gp} \sin^2\gt ,
\label{ourcond}
\eeq
where the infinite factor $\langle \gt \vert \gt \rangle$ has been devided out. 
This result is rather close to that obtained in \cite{Adam}, where the $\gt$-
dependence of the $O(m)$ term has been found to be
\beq
0.742\sin^2\gt + 0.033\cos^2\gt .
\eeq
The reason for a small discrepancy in the values of numerical coefficients 
is that we have been a bit careless in treating the exponential of operators. 
Factorization of $\cos(c\gf_n + \gf_0)$ should be done by the Baker-Campbell-
Hausdorff (BCH) formula taking into account the commutator (\ref{mixedcr}), 
whose r.h.s. is again an operator. The BCH formula thus generates a chain of 
commutators in the exponential representation of $\cos c\gf$ and this leads to 
a small correction of our result (\ref{ourcond}). The details will be given in 
the forthcoming publication \cite{JL}. In any case, the non-canonical 
commutator (\ref{mixedcr}) is crucial for obtaining the correct chiral 
condensate to $O(m)$ in the bosonized massive light-front Schwinger model. 

\section{Acknowledgments}

This work has been supported in part by the Grant No. 2/1156/94 of the Slovak 
Grant Agency for Science, in part by the NSF Grant No. INT-9515511 and by the 
U.S. Department of Energy, Grant No. DE-FG02-87ER40371.

\end{document}